\documentclass[aps,prl,twocolumn,amsmath,amssymb,superscriptaddress,reprint]{revtex4-2}

\usepackage{amsmath,mathtools}
\usepackage{graphicx}
\usepackage[dvipsnames]{color}
\usepackage{hyperref}
\usepackage{psfrag}
\usepackage{stmaryrd}
\usepackage{amssymb}
\usepackage{wasysym}
\usepackage{float}
\usepackage{color}
\usepackage{centernot}

\usepackage[T1]{fontenc}
\usepackage[french,english]{babel}
\usepackage[utf8]{inputenc}
\usepackage{lmodern}

\newcommand{\sgn}{\mathop{\mathrm{sgn}}}

\begin{document}

\title{Energy spectra of non-local internal gravity wave turbulence}

\author{Nicolas Lanchon}
\affiliation{Universit\'e Paris-Saclay, CNRS, FAST, 91405 Orsay, France}
\author{Pierre-Philippe~Cortet}
\email[]{pierre-philippe.cortet@universite-paris-saclay.fr}
\affiliation{Universit\'e Paris-Saclay, CNRS, FAST, 91405 Orsay, France}

\date{\today}

\begin{abstract}
Starting from the classical formulation of the weak turbulence theory in a 
density stratified fluid, we derive a simplified version of the kinetic 
equation of internal gravity wave turbulence. This equation allows us to 
uncover scaling laws for the spatial and temporal energy spectra of internal 
wave turbulence which are consistent with typical scaling exponents observed in 
the oceans. The keystone of our description is the assumption that the energy 
transfers are dominated by a class of non-local resonant interactions, known as 
the ``induced diffusion'' triads, which conserve the ratio between the wave 
frequency and vertical wave number. Our analysis remarkably shows that the 
internal wave turbulence cascade is associated to an apparent constant 
flux of wave action.
\end{abstract}

\maketitle

\textit{Introduction.---} 
Fluids that are stably stratified in density support the propagation of a specific class of waves, called internal gravity waves~\cite{Staquet2002,Sutherland2010,Dauxois2018}. In the case of a linear density gradient, their dispersion relation is
\begin{equation}\label{eq:disp}
\omega = N \frac{k_\perp}{\sqrt{k_\perp^2 + k_z^2}} \, ,
\end{equation}
where $\omega$ is the angular frequency, and $k_\perp$ and $k_z$ are the norm of the components of the wavevector ${\bf k}$ normal and parallel to gravity, respectively. The buoyancy frequency $N=\sqrt{-g/\rho_0\,d\bar{\rho}/dz}$ is set by the density gradient at rest, $d\bar{\rho}/dz<0$, and the acceleration of gravity $g$ (with the vertical coordinate $z$ opposite to gravity). Equation~(\ref{eq:disp}) is obtained from the Navier-Stokes equation under the approximation of weak density variations with respect to the reference density $\rho_0$~\cite{Sutherland2010,Dauxois2018}.

A stratification in density of the fluid, and especially the consequent internal wave dynamics, deeply modifies hydrodynamic turbulence~\cite{Davidson2013,Riley2012}, which becomes anisotropic and can develop in several regimes (see the introduction of Ref.~\cite{Lanchon2023}). A remarkable regime is expected when the Reynolds number $Re= u \ell/\nu$ is large whereas the Froude number $Fr=u/N \ell$ is low compared to the non-dimensional frequency $\omega^*=\omega/N$ (where $u$ and $\omega$ are the characteristic velocity and frequency of the structures at scale $\ell$, respectively, and $\nu$ is the kinematic viscosity). This is the ``weak turbulence'' regime~\cite{Nazarenko2011,Galtier2022}, in which an energy cascade is expected to result from triadic resonant interactions within a statistical ensemble of weakly non-linear internal gravity waves~\cite{Lvov2010}.

This ``weak internal-wave turbulence'' framework has often been suggested as a potential explanation for the oceanic dynamics at ``small scales''~\cite{McComas1981,Polzin2011}, without however a clear confirmation so far. This question is of interest in view of the advance that a validation of the weak turbulence theory could bring for the parameterization of the oceanic ``small scales'' in climate models~\cite{Stensrud2007,Polzin2014,MacKinnon2017,Gregg2018}.

In practice, oceanic data classically reveal one-dimensional (1D) energy spectra, in frequency $\omega$ or in vertical wavenumber $k_z$, following power laws with an exponent of the order of $-2$~\cite{Polzin2011}. These scaling laws are proposed to result from a cascade of energy from low to high frequencies (periods typically in the range from $12$~h to a few tens of minutes) and from large to small vertical scales (typically from a few hundred meters to a few meters). This small-scale high-frequency oceanic behavior is often summarized by the two-dimensional (2D) energy spectrum $E(k_z,\omega) \sim k_z^{-2} \omega^{-2}$ introduced by Garrett and Munk (GM) in the 1970s~\cite{Garrett1972,Garrett1975,Garrett1979} and which postulates a decorrelation between $\omega$ and $k_z$.

Besides, the classical derivation of the wave turbulence theory in stratified fluids, based on the assumption of local interactions in the space of scales, led to an analytical prediction for the 2D (axisymmetric) spatial energy spectrum scaling as $E(k_\perp,k_z) \sim \sqrt{\varepsilon N}\,k_\perp^{-3/2} k_z^{-3/2}$~\cite{Pelinovsky1977,Caillol2000,Lvov2001}, with $\varepsilon$ the mean rate of energy transfer (per unit mass). This derivation has however been achieved by ignoring a divergence of the so-called ``collision integral'' and the relevance of this prediction is therefore highly questionable~\cite{Caillol2000}. Over the past 20~years, several theoretical works have searched to solve this issue by taking into account non-local interactions~\cite{Lvov2010,Dematteis2021}. These works first suggested that a whole family of solutions with a constant energy flux exists~\cite{Lvov2004,Lvov2010} before identifying that the spectrum $E(k_\perp,k_z) \sim k_\perp^{2-a} k_z^{-1}$ with $a\simeq 3.69$ is a remarkable solution because it leads to an exact compensation of two diverging parts of the collision integral~\cite{Lvov2010,Dematteis2021}.

In this Letter, we present a derivation of the 1D energy spectra of weak internal-gravity-wave turbulence based on a detailed analysis of the kinetic equation. The obtained spectra are consistent with the previously mentioned typical oceanic observations. A key step in this derivation is to realize that the turbulent cascade is driven by a subset of triadic resonant interactions, non-local in wavenumber and frequency, which impose a constant ratio between the frequency and the vertical wave number.

\textit{The kinetic equation.---} Starting from the Euler equation under the Boussinesq approximation (in the case of a linear gradient of density at rest), the first step of the weak turbulence theory consists in establishing an evolution equation for the so-called wave action spectrum $n_\mathbf{k}$, a quantity which can be related to the 2D axisymmetric spatial energy spectrum by $E(k_\perp,k_z,t) = k_\perp \omega_\mathbf{k} \, n_\mathbf{k}$~\cite{Caillol2000}. This task was achieved by Caillol and Zeitlin in 2000~\cite{Caillol2000} under the assumptions of weak non-linearity, statistical axisymmetry with respect to gravity and strong anisotropy $k_\perp \ll |k_z|$. Caillol and Zeitlin more precisely established the so-called ``kinetic equation'' for $n_\mathbf{k}$, which reveals the domination of the energy transfers by triadic resonances of internal waves and which can be written as
\begin{equation}\label{eq:cinetic}
    \frac{\partial n_\mathbf{k}}{\partial t} \propto \int \left(\mathcal{R}^{\mathbf{k}}_{\mathbf{pq}} - \mathcal{R}^{\mathbf{p}}_{\mathbf{kq}} - \mathcal{R}^{\mathbf{q}}_{\mathbf{kp}} \right) d\mathbf{p} d \mathbf{q} \, ,
\end{equation}
with
\begin{eqnarray}
    \mathcal{R}^{\mathbf{k}}_{\mathbf{pq}} = T_{\mathbf{kpq}} (n_\mathbf{p} n_\mathbf{q} - n_\mathbf{k} n_\mathbf{p} - n_\mathbf{k} n_\mathbf{q}) \delta^\mathbf{k}_\mathbf{pq} \delta(\Omega^\mathbf{k}_\mathbf{pq}) \, , \label{eq:R}\\
    \begin{split} \label{eq:coefTkpq}
        T_{\mathbf{kpq}} = \left( \tilde{k}_\perp + \tilde{p}_\perp + \tilde{q}_\perp \right)^2 \frac{\left(k_z^2-p_z q_z \right)^2}{16 |k_z p_z q_z| k_\perp p_\perp q_\perp}  \\  \times\left( \frac{k_\perp^2 - \tilde{p}_\perp \tilde{q}_\perp}{k_z^2-p_z q_z}k_z - \frac{p_\perp^2}{p_z} - \frac{q_\perp^2}{q_z} \right)^2 \, ,
    \end{split}
\end{eqnarray}
and $\tilde{m}_\perp = \sgn(m_z) m_\perp$ (with $\mathbf{m}=\mathbf{k},\mathbf{p}$ or $\mathbf{q}$), $\delta^\mathbf{k}_\mathbf{pq} = \delta(\mathbf{k} - \mathbf{p} - \mathbf{q})$ and $ \Omega^\mathbf{k}_\mathbf{pq} = \omega_\mathbf{k}^* - \omega_\mathbf{p}^* - \omega_\mathbf{q}^*$ (see Ref.~\cite{Lvov2012} for a review on the internal wave kinetic equation). The non-dimensional angular frequencies $\omega_\mathbf{m}^*$ (with $\mathbf{m}=\mathbf{k},\mathbf{p}$ or $\mathbf{q}$) verify the dispersion relation~(\ref{eq:disp}) which, in the considered anisotropic limit, reduces to $\omega_\mathbf{m}^* = m_\perp/|m_z|$.

At this step, the usual strategy to find a physical solution consists in searching for a stationary solution of the kinetic equation~(\ref{eq:cinetic}) with a non-zero energy flux. To achieve this, Caillol and Zeitlin~\cite{Caillol2000} employed the Zakharov-Kuznetsov transformation~\cite{Nazarenko2011} in the right-hand side (rhs) of Eq.~(\ref{eq:cinetic}) and finally identified the stationary solution mentioned earlier
\begin{equation}\label{eq:localSolution}
E(k_\perp,k_z) \sim \sqrt{\varepsilon N} k_\perp^{-3/2} k_z^{-3/2} \,.
\end{equation}

In this paragraph, we explain how Eq.~(\ref{eq:localSolution}) can also be derived from the kinetic equation~(\ref{eq:cinetic}) using phenomenological arguments.
A key point is to assume the locality of interactions both in wave number ($|\mathbf{k}|\sim |\mathbf{p}|\sim |\mathbf{q}|$) and frequency ($\omega_\mathbf{k}^* \sim \omega_\mathbf{p}^* \sim \omega_\mathbf{q}^*$), which amounts to consider that $k_\perp \sim p_\perp \sim q_\perp$ and $|k_z| \sim |p_z| \sim |q_z|$. Then, analyzing the scaling of the transfer coefficients in Eqs.~(\ref{eq:cinetic}) to (\ref{eq:coefTkpq}), we can estimate the transfer time $\tau_{tr}$ as
\begin{equation}\label{eq:tau}
    \tau_{tr} \sim \frac{\omega_\mathbf{k}^*}{T_\mathbf{kpq} n_\mathbf{k} k_\perp^2 k_z} \sim \frac{\omega_\mathbf{k}^*}{n_\mathbf{k} k_\perp^5} \sim \frac{\omega_\mathbf{k}}{k_\perp^2 u_\mathbf{\perp}^2} \, ,
\end{equation}
where $T_\mathbf{kpq} \sim k_\perp^3/k_z$ and $u_\perp$ represents the typical velocity at scale $\mathbf{k}$. We used here the estimate $n_\mathbf{k} \sim u_\perp^2/k_\perp^3 N$ resulting from the definition of the wave action spectrum $n_\mathbf{k} = E(k_\perp,k_z) / k_\perp \omega_\mathbf{k}$ and the estimate of the power spectral density $E(k_\perp,k_z)$ using $u_\perp^2 \sim E(k_\perp,k_z) k_\perp k_z$. For the last scaling law, we considered the fact that the kinetic and potential energies of an internal gravity wave are equal. Let us also highlight that, under the considered anisotropic assumption $k_\perp \ll |k_z|$, the kinetic energy is dominated by the horizontal component of the velocity which explains our notation $u_\perp$. From Eq.~(\ref{eq:tau}), one can finally recover the energy spectrum~(\ref{eq:localSolution}) considering that the transfer time $\tau_{tr}$ should also verify $\varepsilon \sim u_{\perp}^2/\tau_{tr}$. Furthermore, introducing the non-linear time $\tau_{nl} \sim 1/k_\perp u_\perp$~\cite{footnote1}, we can note that the obtained transfer time scales as $\tau_{tr} \sim \omega_\mathbf{k} \tau_{nl}^2$. This scaling is classical of wave turbulence systems where energy transfers are governed by triadic and local wave interactions~\cite{Galtier2022}.

As mentioned in the introduction and regardless of the beauty of this result, injecting a posteriori the solution~(\ref{eq:localSolution}) in the collision integral, i.e. the right-hand side of Eq.~(\ref{eq:cinetic}), leads to a divergence~\cite{Caillol2000}. This renders the solution~(\ref{eq:localSolution}) unacceptable. This ``failure'' of the Zakharov transformation illustrates the fact that it is relevant only when interactions local in wavenumbers and frequencies are dominant. It actually suggests that ``non-local'' triadic interactions are most probably driving the turbulent dynamics.

\textit{Non-locality.---} To support this idea, we can evaluate the transfer coefficients $T_{\mathbf{kpq}}$, $T_{\mathbf{pkq}}$, and $T_{\mathbf{qpk}}$ of the collision integral. For each of them, the analysis must be conducted on the resonant manifold, defined by $\mathbf{k} = \mathbf{p} + \mathbf{q}$ and $\omega_\mathbf{k}^* = \omega_\mathbf{p}^* + \omega_\mathbf{q}^*$ for $T_{\mathbf{kpq}}$ and their relevant permutations for the two other coefficients, in line with the delta functions in Eq.~(\ref{eq:R}). The coefficients $T_{\mathbf{kpq}}$, $T_{\mathbf{pkq}}$, and $T_{\mathbf{qpk}}$ are thus 9-variable functions constrained by four resonance equations. Further fixing the wave vector $\mathbf{q}$, the resonance manifolds can be parameterized by two variables only and the coefficient $T_{\mathbf{kpq}}$ (or its permutations) mapped on these two variables (we will choose $\omega_\mathbf{p}^*=p_\perp/|p_z|$ and $p_z$). Without loss of generality, we choose $q_z > 0$, and, since there is no viscosity, we can also take $q_z = 1$. Furthermore, due to axial symmetry, we can choose $q_y = 0$ and $q_x > 0$. Finally, we choose a low frequency $\omega_\mathbf{q}^* = 0.001$, which sets $q_x = 0.001$, in order to fulfill the strong anisotropy condition.

\begin{figure}
    \centering
    \includegraphics[width=8.5cm]{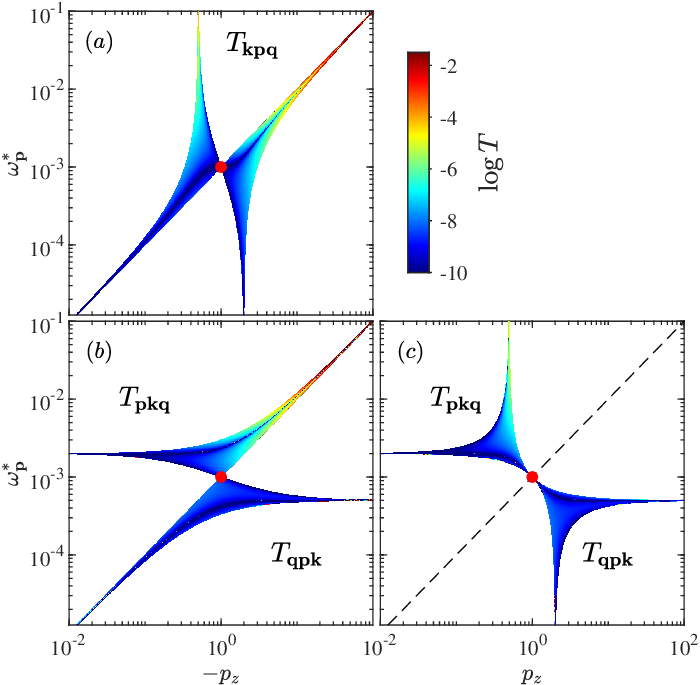}
    \caption{Logarithm of the coefficients $T_\mathbf{kpq}$, $T_\mathbf{pkq}$, and $T_\mathbf{qpk}$ evaluated on their respective resonant manifold for $\mathbf{q} = (0.001,0,1)$. 
    The x-axis shows $-p_z$ in panels (a) and (b) and $p_z$ in panel (c). The red point highlights the point $(|p_z|,\omega_\mathbf{p}^*)=(|q_z|,\omega_\mathbf{q}^*)$ and the dashed lines correspond to the condition $\omega_\mathbf{p}^*/|p_z| = \omega_\mathbf{q}^*/|q_z|$.}
    \label{fig:CoefInteraction}
\end{figure}

Thus, Fig.~\ref{fig:CoefInteraction} shows the logarithm of the coefficients $T_\mathbf{kpq}$, $T_\mathbf{pkq}$, and $T_\mathbf{qpk}$ evaluated on their respective resonant manifold for $\mathbf{q} = (0.001,0,1)$. A careful analysis of the different panels reveals that two specific branches of the resonant manifolds are associated to transfer coefficients several orders of magnitude larger than everywhere else. These resonances are found in panels (a) and (b) (for $T_\mathbf{kpq}$ and $T_\mathbf{pkq}$, respectively) along the line of equation $\omega_\mathbf{p}^*/|p_z| = \omega_\mathbf{q}^*/|q_z|$ and for $\omega_\mathbf{p}^*$ and $|p_z|$ much larger than $\omega_\mathbf{q}^*$ and $|q_z|$, respectively. 

These observations suggest that the energy transfers of a strongly-anisotropic weakly-nonlinear stratified turbulence are mediated by resonant wave triads verifying 
\begin{eqnarray}
    \omega_\mathbf{q}^* &\ll \omega_\mathbf{p}^* \sim \omega_\mathbf{k}^* \,, \label{eq:ID1}\\
    |\mathbf{q}| &\ll |\mathbf{p}| \sim |\mathbf{k}|\,. \label{eq:ID2}
\end{eqnarray}
In the literature, the resonant triads fulfilling these conditions are referred to as the ``induced diffusion'' triads~\cite{McComas1977}. Their domination over the energy transfers in the weak internal-wave turbulence regime has already been evidenced by previous analyses~\cite{Lvov2010,Dematteis2021}.

Another key feature that is specific to the ``induced diffusion'' triads is the conservation of the ratio 
\begin{equation}\label{eq:ID3}
    \frac{\omega_\mathbf{q}^*}{|q_z|} = \frac{\omega_\mathbf{p}^*}{|p_z|} =  \frac{\omega_\mathbf{k}^*}{|k_z|}\, .
\end{equation}
This feature can be seen in Fig.~\ref{fig:CoefInteraction}, but Eq.~(\ref{eq:ID3}) can actually also be demonstrated from Eqs.~(\ref{eq:ID1}) and (\ref{eq:ID2}), the wave dispersion relation and the resonance conditions [see Supplemental Material (SM)]. This property will be central in the description of the internal wave turbulence that we propose in the following.

\textit{A description of the ``induced diffusion'' wave turbulence.---} We proceed by assuming that internal-wave turbulence is driven only by triadic resonant interactions verifying Eqs.~(\ref{eq:ID1}) and (\ref{eq:ID2}). We introduce the characteristic length $\xi$, such that $\omega_\mathbf{k}^* = \xi |k_z|$ and $k_\perp = \xi k_z^2$, which is expected to be conserved along the turbulent cascade.

Thanks to this ``induced diffusion'' assumption, we can simplify the kinetic equation [i.e., Eqs.~(\ref{eq:cinetic}-\ref{eq:coefTkpq})]. First, we can show that the coefficients $T_{\mathbf{kpq}}$ and $T_{\mathbf{pkq}}$ (when evaluated on their respective resonant manifolds) can be written as
\begin{equation}\label{eq:TID}
    T = T_{\mathbf{kpq}} = T_{\mathbf{pkq}} = \frac{q_\perp}{|q_z|} k_\perp^2 \cos^2(\varphi_{\mathbf{kq}}) \, ,
\end{equation}
where $\varphi_{\mathbf{kq}}$ denotes the angle between the projections of $\mathbf{k}$ and $\mathbf{q}$ in the horizontal plane (see SM for the demonstration). 
We remark that the coefficient $T$ vanishes when the projections of $\mathbf{k}$ and $\mathbf{q}$ in the horizontal plane are orthogonal, i.e., when $\varphi_{\mathbf{kq}} = \pm \pi/2$. This cancellation is visible in Fig.~\ref{fig:CoefInteraction}, where we see on panels (a) and (b) a dark blue ``cancellation'' line in the middle of the induced diffusion branches. On the other hand, the transfer coefficient (\ref{eq:TID}) is maximized when $\mathbf{k}$, $\mathbf{p}$, and $\mathbf{q}$ lie in the same vertical plane.

Using methods inspired by Refs.~\cite{Balk1990,Nazarenko2001,Galtier2022} and horizontal isotropy, we can further show that the kinetic equation simplifies to
\begin{equation}
    \frac{\partial n_\mathbf{k}}{\partial t} \propto \frac{3}{4k_\perp} \frac{\partial}{\partial k_\perp} k_\perp D_{\perp}(\mathbf{k})  \frac{\partial n_\mathbf{k}}{\partial k_\perp} + \frac{\partial}{\partial k_z} D_{z}(\mathbf{k})  \frac{\partial n_\mathbf{k}}{\partial k_z} \, , \label{eq:eqkineticID}
\end{equation}
with 
\begin{equation}\label{eq:kineticID2}
    D_{i}(\mathbf{k}) = \int q_\perp^2 q_i^2 k_\perp^2 n_\mathbf{q} {\delta}{\left( q_\perp -  q_z^2 \frac{k_\perp}{k_z^2} \right)} d q_\perp d q_z \, ,
\end{equation}
where $i = \, \perp$ or $i = \, z$ (see details in SM). Following from the assumption that the energy transfers are non-local, controlled by induced-diffusion triads, the integration in Eq.~(\ref{eq:kineticID2}) is restricted to a ``large scale'' domain $|\mathbf{q}| \leq \tilde{q}$ defined by an arbitrary cutoff wavenumber $\tilde{q}$ much smaller than the norm of wave vector $\mathbf{k}$. One should note that McComas and Bretherton obtained a similar equation in cartesian coordinates in Ref.~\cite{McComas1977}. It is also worth to remark that the ``induced diffusion'' relation $k_\perp/k_z^2 \simeq q_\perp/q_z^2$, equivalent to (\ref{eq:ID3}), naturally emerges here from Eq.~(\ref{eq:eqkineticID}) through the Dirac delta function in~Eq.~(\ref{eq:kineticID2}). 

We then search for a power law steady solution to this equation of the form $n_\mathbf{k} \propto k_\perp^\alpha k_z^\beta$. Introducing this ansatz in Eq.~(\ref{eq:eqkineticID}) [see SM], we find that the only couple of exponents canceling the rhs of the equation leads to a wave action spectrum scaling as  
\begin{equation}
    n_\mathbf{k} \sim k_\perp^{-3} k_z^{-1}\,. \label{eq:solutionNk}
\end{equation}
This scaling law corresponds to a 2D axisymmetric spatial energy spectrum following $E(k_\perp,k_z) \sim k_\perp^{-1} k_z^{-2}$.

To derive Eq.~(\ref{eq:eqkineticID}), we split in the Fourier space the 
flow in two subsystems, separated 
by an arbitrary wavenumber $\tilde{q}$ and which are exchanging energy via 
non-local 
induced-diffusion triads. It is important to note that 
Eq.~(\ref{eq:eqkineticID}), which describes the dynamics of the small-scale 
subsystem, has the structure of a diffusion equation for the wave action 
spectrum $n_\mathbf{k}$. In Eq.~(\ref{eq:eqkineticID}), the transfers of energy 
of the small-scale subsystem with the large scales are accounted for by the 
effective diffusion coefficients~(\ref{eq:kineticID2}) 
which are dependent on the amplitude of the large-scale modes.
It is this property that actually led to the name of ``induced-diffusion'' 
triads. This type of 
diffusion equations conserves the wave action of the small-scale subsystem as 
noticed by McComas and Bretherton~\cite{McComas1977} as well 
as by Nazarenko~\cite{Nazarenko2011} in the description of non-local Rossby 
drift wave turbulence. This conservation implies that the turbulent 
cascade described by Eq.~(\ref{eq:eqkineticID}) is associated to an apparent 
constant local flux of wave action $\zeta$ and to an apparent local flux of 
energy $\varepsilon_{\rm app} \equiv \zeta \omega_\mathbf{k} \sim \zeta N 
\xi k_z$ which is not constant. These features are in line with the fact the 
energy of the small-scale subsystem described by Eq.~(\ref{eq:eqkineticID}) is 
not conserved. There is however no contradiction with the conservation of the 
energy of the whole system. Indeed, because of the 
non-locality of the energy transfers, the large-scale subsystem 
behaves as a source of energy at each scale of the small-scale 
subsystem.

To identify the prefactor of the spectrum~(\ref{eq:solutionNk}), we write 
the scaling law expected for the diffusive time $\tau_{d}$ from the 
analysis of Eqs.~(\ref{eq:eqkineticID}) and (\ref{eq:kineticID2}). This leads 
to the relation $1/\tau_{d} \sim N \Psi k_\perp^2/k_z^2$ where $\Psi$ is 
a non-dimensional number defined by the integral
\begin{equation}
    \Psi = \int \frac{q_\perp^2 q_z^2}{N} n_\mathbf{q} {\delta}{\left(q_\perp - \xi q_z^2\right)} d q_\perp d q_z \label{eq:defPsi}
\end{equation}
over the ``large scale'' domain ($|\mathbf{q}| \leq \tilde{q}$) and where $\xi 
= k_\perp/k_z^2$. Let us note that we considered the second term of the rhs of 
Eq.~(\ref{eq:eqkineticID}) since using the first term would lead to a time 
larger by a factor of $\omega_\mathbf{k}^2/\omega_\mathbf{q}^2$. Using $\xi = 
k_\perp/k_z^2$ and the scaling relation $\zeta \sim n_\mathbf{k} k_\perp^2 
k_z/\tau_{d}$ between the apparent flux of wave action $\zeta$ and the wave 
action spectrum $n_\mathbf{k}$, we find a scaling law in line with 
Eq.~(\ref{eq:solutionNk})
\begin{equation}
    n_\mathbf{k} \sim \frac{\zeta}{\Psi \xi N} k_\perp^{-3} k_z^{-1} \, . \label{eq:spectre2D}
\end{equation}

At this step, we note that since the cutoff wave number $\tilde{q}$ can be chosen arbitrarily, Eq.~(\ref{eq:spectre2D}) should remain valid for all wave vectors $\mathbf{k}$, with $\Psi$ being a constant. We can then inject Eq.~(\ref{eq:spectre2D}) in Eq.~(\ref{eq:defPsi}) and find an estimate of $\Psi$ in terms of $\zeta$ and $\xi$ only: $\Psi^2 \sim \zeta/(N^2 \xi^2)$.
Using the relation $E(k_\perp,k_z) = k_\perp \omega_\mathbf{k} \, 
n_\mathbf{k}$, we finally obtain a comprehensive scaling law for the 2D spatial 
energy spectrum
\begin{equation}
    E(k_\perp,k_z) \sim \sqrt{\zeta} N k_\perp^{-1} k_z^{-2} \, . \label{eq:spectre2Dbis}
\end{equation}

Following this result, we obtain a scaling law for the 1D ``vertical'' spatial energy spectrum $E(k_z) \sim E(k_\perp,k_z) k_\perp$ of the form 
\begin{equation}
    E(k_z) \sim \sqrt{\zeta} N k_z^{-2} \, . \label{eq:spectreTurbOndesKz}
\end{equation}
Furthermore, using the correspondences between the scaling law of the different 1D energy spectra, $E(k_z) k_z \sim E(k_\perp) k_\perp \sim E(\omega) \omega$, coupled to the ``induced diffusion'' relation $k_\perp = \xi k_z^2$, we obtain the scaling laws of the ``horizontal'' and ``temporal'' 1D energy spectra
\begin{align}
E(k_\perp) &\sim \sqrt{\zeta \xi} N k_\perp^{-3/2} \, , \label{eq:spectreTurbOndesKperp} \\
E(\omega) &\sim \sqrt{\zeta} \xi N^2 \, \omega^{-2} \, . \label{eq:spectreTurbOndesOmega}
\end{align}
It is remarkable that the exponents reported for the 1D energy spectra, as a function of $k_z$ in Eq.~(\ref{eq:spectreTurbOndesKz}) and as a function of $\omega$ in Eq.~(\ref{eq:spectreTurbOndesOmega}), are compatible with classical \textit{in situ} observations in the oceans~\cite{Polzin2011}. We should also recall that, since for an internal wave, the kinetic and potential energies are equal, the spectra~(\ref{eq:spectreTurbOndesKz}-\ref{eq:spectreTurbOndesOmega}) can be understood as kinetic, potential or total energy spectra.

\textit{Beyond wave turbulence.---} The primary assumption of the wave turbulence theory is weak nonlinearity, meaning that the non-linear time $\tau_{nl}$ is much larger than the wave period, i.e. $\omega_\mathbf{k} \tau_{nl} \gg 1$. Using the scaling $\tau_{nl} \sim 1/k_\perp u_\perp$ of the non-linear time in the anisotropic limit $k_\perp \ll |k_z|$~\cite{footnote1}, Eq.~(\ref{eq:spectreTurbOndesKperp}) leads to the scaling $\omega_\mathbf{k} \tau_{nl}\sim N^{1/2} \xi^{1/4} \zeta^{-1/4}\,k_\perp^{-1/4}$ for the non-linearity parameter. To derive this relation, we used the fact the length $\xi=\omega_\mathbf{k}^*/|k_z|=k_\perp/k_z^2$ is conserved under the ``induced diffusion'' assumption. This scaling implies that the turbulent cascade will depart from the weak non-linearity condition beyond the cutoff wave vector 
\begin{equation}
    (\kappa_\perp,\kappa_z) = \left(\frac{N^2 \xi}{\zeta},\frac{N}{\sqrt{\zeta}}\right) = \left(\frac{N^6 \xi^3}{\epsilon^2},\frac{N^3 \xi}{\epsilon} \right) \,.
\end{equation}
At larger wave vectors, the turbulence is expected to enter the so-called 
``strongly stratified turbulence'' regime~\cite{Billant2001,Lindborg2006,Brethouwer2007,Nazarenko2011a}. This 
regime involves local triadic and strongly non-linear interactions driving a 
constant energy flux~$\epsilon$ (equal to $\varepsilon_{\rm app} \sim 
\zeta \omega_\mathbf{k}$ at the cross-over scale). It leads to vertical and 
horizontal 1D kinetic energy spectra scaling as $E(k_z) \sim N^2 k_z^{-3}$ and 
$E(k_\perp) \sim \epsilon^{2/3} k_\perp^{-5/3}$, respectively. Remarkably, 
evidences of such a transition from weakly to strongly non-linear stratified 
turbulence have been reported at small oceanic scales in the literature (see, 
e.g., Fig.~1 in Ref.~\cite{Riley2008} and Fig.~21 in Ref.~\cite{Polzin2011}).

\textit{Conclusion.---} In this Letter, we derive scaling laws for the energy spectra of internal gravity wave turbulence. We start from the kinetic equation obtained under the assumptions of weak non-linearity, statistical axisymmetry and strong anisotropy $k_\perp \ll |k_z|$~\cite{Caillol2000}. Following previous works~\cite{Lvov2010,Dematteis2021}, our numerical analysis of the collision integral suggests that the energy transfers are dominated by a specific class of non-local triadic resonant interactions which are referred to as the ``induced diffusion'' triads in the literature~\cite{McComas1977}. We further show that these triads have the remarkable property of keeping constant the ratio between the wave frequency $\omega_\mathbf{k}$ and the vertical wave number $|k_z|$ defining a conserved characteristic length $\xi=\omega_\mathbf{k}/N|k_z|$ ($N$ is the buoyancy frequency). It is worth to note that this feature departs from an assumption of the Garrett and Munk model~\cite{Garrett1972,Garrett1975,Garrett1979} for a finite depth ocean, which is the decorrelation between the frequency and vertical wave number.

Building on these results, we derive analytically a simplified version of the kinetic equation assuming that only the ``induced diffusion'' triads contribute to the internal wave turbulent cascade. This kinetic equation has the structure of a diffusion equation for the wave action spectrum which results from the scale separation within the ``induced diffusion'' triads. We show that this kinetic equation has only one power-law steady solution, which corresponds to a 2D axisymmetric spatial energy spectrum following $E(k_\perp,k_z) \sim k_\perp^{-1} k_z^{-2}$. This scaling further leads to 1D energy spectra with an exponent $-2$, both as a function of the frequency and of the vertical wave number, which feature is in line with classical scaling exponents observed in the oceans~\cite{Polzin2011}. In parallel, a scaling $E(k_\perp) \sim k_\perp^{-3/2}$ is predicted for the 1D horizontal spatial energy spectrum.

A complementary dimensional analysis of the simplified kinetic equation allows 
us to identify the prefactors of the energy spectra. Our analysis remarkably 
shows that, following from 
the non-locality of the energy transfers, the internal wave 
turbulence cascade is associated to an apparent constant flux of wave 
action.

Owing to the importance of the assumption that we made in our analytical calculations that only ``induced diffusion'' triads contribute to the internal wave turbulence, our approach and predictions are to be validated by alternative strategies which might be numerical simulations or experiments of a genuine weakly non-linear internal wave turbulence. Beyond that, it will be crucial to assess the relevance of this approach to describe the small-scale oceanic dynamics, since major advances for the parameterization of the oceanic ``small scales'' in global climate models might be expected.

\begin{acknowledgments}
This work was supported by a grant from the Simons Foundation (651461, PPC). 
\end{acknowledgments}


\begin{thebibliography}{99}

\bibitem{Staquet2002} C. Staquet and J. Sommeria, Internal gravity waves: From instabilities to turbulence, Annu. Rev. Fluid Mech. \textbf{34}, 559 (2002).

\bibitem{Sutherland2010} B.R. Sutherland, \textit{Internal Gravity Waves} (Cambridge University Press, Cambridge, UK, 2010).

\bibitem{Dauxois2018} T. Dauxois, S. Joubaud, P. Odier, and A. Venaille, Instabilities of internal wave beams, Annu. Rev. Fluid Mech. \textbf{50}, 131 (2018).

\bibitem{Davidson2013} P.A. Davidson, \textit{Turbulence in Rotating, Stratified and Electrically Conducting Fluids} (Cambridge University Press, Cambridge, UK, 2013).

\bibitem{Riley2012} J. Riley and E. Lindborg, Recent Progress in Stratified Turbulence, in \textit{Ten Chapters in Turbulence} edited by P. Davidson, Y. Kaneda and K. Sreenivasan, 269 (Cambridge University Press, Cambdrige, UK, 2012). 

\bibitem{Lanchon2023} N. Lanchon, D.O. Mora, E. Monsalve, and P.-P. Cortet, Internal wave turbulence in a stratified fluid with and without eigenmodes of the experimental domain, Phys. Rev. Fluids \textbf{8}, 054802 (2023).

\bibitem{Nazarenko2011} S. Nazarenko, {\it Wave  Turbulence} (Springer, Berlin, 2011).

\bibitem{Galtier2022} S. Galtier, \textit{Physics of Wave Turbulence} (Cambridge University Press, Cambridge, UK, 2022).

\bibitem{Lvov2010} Y.V. Lvov, K.L. Polzin, E.G. Tabak, and N. Yokoyama, Oceanic internal-wave field: Theory of scale-invariant spectra, J. Phys. Oceanogr. \textbf{40}, 2605 (2010).

\bibitem{Polzin2011}  K.L. Polzin and Y.V. Lvov, Toward regional characterizations of the oceanic internal wavefield, Rev. Geophys. \textbf{49}, RG4003 (2011).

\bibitem{McComas1981}  C.H. McComas and P. Müller, The dynamic balance of internal waves, J. Phys. Oceanogr. \textbf{11}, 970 (1981).

\bibitem{Stensrud2007} D.J. Stensrud, \textit{Parametrization Schemes: Keys to Understanding Numerical Weather Prediction Models} (Cambridge University Press, Cambridge, UK, 2007).

\bibitem{Polzin2014} K.L. Polzin, A.C.N. Garabato, T.N. Huussen, B.M. Sloyan, and S. Waterman, Finescale parameterizations of turbulent dissipation, J. Geophys. Res. Oceans \textbf{119}, 1383 (2014).

\bibitem{MacKinnon2017} J.A. MacKinnon, Z. Zhao, C.B. Whalen, A.F. Waterhouse, D.S. Trossman, O.M. Sun, L.C. St. Laurent, H.L. Simmons, K. Polzin, R. Pinkel, A. Pickering, N.J. Norton, J.D. Nash, R. Musgrave, L.M. Merchant, A.V. Melet, B. Mater, S. Legg, W.G. Large, E. Kunze, J.M. Klymak, M. Jochum, S.R. Jayne, R.W. Hallberg, S.M. Griffies, S. Diggs, G. Danabasoglu, E.P. Chassignet, M.C. Buijsman, F.O. Bryan, B.P. Briegleb, A. Barna, B.K. Arbic, J.K. Ansong, and M.H. Alford, Climate Process Team on Internal Wave-Driven Ocean Mixing, Bull. Am. Meteorol. Soc. \textbf{98}, 2429 (2017).

\bibitem{Gregg2018} M.C. Gregg, E.A. D'Asaro, J.J. Riley, and E. Kunze, Mixing efficiency in the ocean, Annu. Rev. Mar. Sci. \textbf{10}, 443 (2018).

\bibitem{Garrett1972} C. Garrett and W. Munk, Space-Time Scales of Internal Waves, Geophys. Fluid Dyn. \textbf{2}, 225 (1972).

\bibitem{Garrett1975} C. Garrett and W. Munk, Space-Time Scales of Internal Waves: A Progress report, J. Geophys. Res. \textbf{80}, 291 (1975).

\bibitem{Garrett1979} C. Garrett and W. Munk, Internal waves in the ocean, Annu. Rev. Fluid. Mech. \textbf{11}, 339 (1979).

\bibitem{Pelinovsky1977} E.N. Pelinovsky and M.A. Raevsky, Weak turbulence of 
internal waves in the ocean, Atm. Ocean Phys. Izvestija \textbf{13}, 187 (1977).

\bibitem{Caillol2000} P. Caillol and V. Zeitlin, Kinetic equations and stationary energy spectra of weakly nonlinear internal gravity waves, Dyn. Atmos. Oceans \textbf{32}, 81 (2000).

\bibitem{Lvov2001} Y.V. Lvov and E.G. Tabak, Hamiltonian Formalism and the Garrett-Munk Spectrum of Internal Waves in the Ocean, Phys. Rev. Lett. \textbf{87}, 168501 (2001).

\bibitem{Dematteis2021} G. Dematteis and Y.V. Lvov, Downscale energy fluxes in scale-invariant oceanic internal wave turbulence, J. Fluid Mech. \textbf{915}, 1 (2021).

\bibitem{Lvov2004} Y.V. Lvov, K.L. Polzin, and E.G. Tabak, Energy Spectra of the Ocean’s Internal Wave Field: Theory and Observations, Phys. Rev. Lett. \textbf{92}, 128501 (2004). 

\bibitem{Lvov2012} Y.V. Lvov, K.L. Polzin and N. Yokoyama, Resonant and Near-Resonant Internal Wave Interactions, J. Phys. Oceanogr. \textbf{42}, 669 (2012).

\bibitem{footnote1} The scaling $\tau_{nl} \sim 1/k_\perp u_\perp$ for the non-linear time follows from the expression of the wave triadic interaction coefficients in the anisotropic limit $k_\perp \ll |k_z|$. The expression of the triadic interaction coefficients can be found in Ref.~\cite{Remmel2014}.

\bibitem{McComas1977} C.H. McComas and F.P. Bretherton, Resonant Interaction of Oceanic Internal Waves, J. Geophys. Res. \textbf{82}, 9 (1977).

\bibitem{Balk1990} A.M. Balk, S.V. Nazarenko and  V.E. Zakharov, On the nonlocal turbulence of drift type waves, Phys. Lett. A \textbf{146}, 217 (1990).

\bibitem{Nazarenko2001} S.V. Nazarenko, A.C. Newell and S. Galtier, Non-local MHD turbulence, Physica D: Nonlinear Phenomena, \textbf{152-153}, 646 (2001).

\bibitem{Lindborg2006} E. Lindborg, The energy cascade in a strongly stratified fluid, J. Fluid Mech. \textbf{550}, 207 (2006).

\bibitem{Brethouwer2007} G. Brethouwer, P. Billant, E. Lindborg, and J.-M. Chomaz, Scaling analysis and simulation of strongly stratified turbulent flows, J. Fluid Mech. \textbf{585}, 343 (2007).

\bibitem{Nazarenko2011a} S.V. Nazarenko and A.A. Chekochihin, Critical balance in magnetohydrodynamic, rotating and stratified turbulence: towards a universal scaling conjecture, J. Fluid Mech. \textbf{677}, 134 (2011).

\bibitem{Billant2001} P. Billant and J.-M. Chomaz, Self-similarity of strongly stratified inviscid flows, Phys. of Fluids \textbf{13}, 1645 (2001).

\bibitem{Riley2008} J.J. Riley and E. Lindborg, Stratified Turbulence: A Possible Interpretation of Some Geophysical Turbulence Measurements, J. Atmos. Sci. \textbf{65}, 2416 (2008).

\bibitem{Remmel2014} M. Remmel, J. Sukhatme, and L.M. Smith, Nonlinear gravity-wave interactions in stratified turbulence, Theor. Comput. Fluid Dyn. \textbf{28}, 131 (2014).

\end{thebibliography}
\end{document}